\documentclass[sigconf]{acmart}

\usepackage{graphicx}
\usepackage{caption}
\usepackage{subcaption}
\usepackage{tabularx}
\usepackage{makecell}
\usepackage{hyperref}
\usepackage{pifont}
\usepackage{booktabs}

\begin{document}

\title{Speculating for Epiplexity: How to Learn the Most from Speculative Design? }

\author{Botao Amber Hu}
\orcid{0000-0002-4504-0941}
\affiliation{%
  \institution{University of Oxford}
  \city{Oxford}
  \country{UK}
  }
\email{botao.hu@cs.ox.ac.uk}

\begin{abstract}
  Speculative design uses provocative "what if?" scenarios to explore possible sociotechnical futures, yet lacks rigorous criteria for assessing the quality of speculation. We address this gap by reframing speculative design through an information-theoretic lens as a resource-bounded knowledge generation process that uses provotypes to strategically embrace surprise. However, not all surprises are equally informative—some yield genuine insight while others remain aesthetic shock. Drawing on epiplexity—structured, learnable information extractable by bounded observers—we propose decomposing the knowledge generated by speculative artifacts into structured epistemic information (transferable implications about futures) and entropic noise (narrative, aesthetics, and surface-level surprise). We conclude by introducing a practical audit framework with a self-assessment questionnaire that enables designers to evaluate whether their speculations yield rich, high-epiplexity insights or remain at a superficial level. We discuss implications for peer review, design pedagogy, and policy-oriented futuring.

\end{abstract}



\maketitle


\section{Introduction}
Speculative design has become a prominent approach for ``what if'' inquiry in HCI, DIS, and adjacent design research communities, aiming to provoke reflection on sociotechnical trajectories rather than to optimize usability or market fit \cite{dunneRaby2013speculativeEverything,wongKhovanskaya2018speculativeHCI,cordova2025slrSpeculative}. Design artifacts---props, prototypes, scenarios, enacted experiences---operate as epistemic devices: they externalize assumptions, materialize futures, and support discursive engagement around values, power, and consequence \cite{sengers2005reflectiveDesign,pierceEtAl2015designCriticality,candyDunagan2017experientialFutures}. This orientation is increasingly relevant for AI and data-intensive systems, where second-order effects---labor displacement, governance drift, contestability failures, racialized harm---are difficult to surface through conventional development pipelines \cite{crawfordJoler2018anatomyAISystem,noble2018algorithmsOppression}. Behavioral scientists have independently converged on this challenge: Rahwan, Shariff, and Bonnefon propose a ``science fiction science'' method that applies controlled experiments to speculative futures, demonstrating growing cross-disciplinary recognition that rigorous approaches to anticipatory inquiry are needed \cite{rahwan2025sciFiSci}.

However, the very qualities that make speculative design generative---openness, ambiguity, friction, plural futures \cite{gaver2003ambiguity}---also complicate how the field judges rigor and contribution. In HCI, speculative and critical work is often read ``in tension with progression,'' where progressional design converges toward implementation while frictional design resists that vector to open interpretive space \cite{pierce2021tensionProgression}. Recent synthesis work makes the evaluative problem explicit: a scoping review of speculative design quality finds criteria dispersed across heterogeneous traditions and proposes a taxonomy of qualities, but underscores that ``quality'' remains hard to stabilize \cite{ringfortfelner2025quality}. A systematic literature review of speculative design processes identifies recurring phases---\emph{select}, \emph{explore}, \emph{transform}, \emph{provoke}---and proposes an ``inverted double diamond'' framework \cite{cordova2025slrSpeculative}. Together, these syntheses clarify both the spread and the methodological indeterminacy of speculative design: we have rich practices, but incomplete shared theory for what makes a provocation \emph{learnable}.

This paper offers an information-theoretic reframing of that problem. We ask: \emph{How can we learn most from provocation?} The urgency is sharpened by what Collingridge termed the \emph{dilemma of control}: the point of maximum leverage over a technology's social trajectory coincides with minimum knowledge of its consequences, while by the time consequences become clear, the technology is entrenched and resistant to change \cite{collingridge1980socialControl,rahwan2025sciFiSci}. We propose that speculative design can be treated as a bounded information process: it constructs partial models of futures and elicits interpretations from bounded observers under constraints of time, attention, and feasibility \cite{simon1955boundedRationality,finzi2026epiplexity}. Within those bounds, not all information is equally useful. Some elements are structured and reusable; others are entropic noise. We call the structured, learnable component \emph{epiplexity}, adapting a recent information-theoretic construct that formalizes ``useful information'' for computationally bounded intelligence \cite{finzi2026epiplexity}.

We contribute:
\begin{enumerate}
\item A \textbf{theoretical model} of speculative design as a bounded information process, contrasting progressional (negentropic) and frictional (strategically entropic) design logics, and decomposing what observers learn into structured information ($S_t$) and residual entropy ($H_t$).
\item A \textbf{four-quadrant diagnostic map} distinguishing qualitatively different outcomes of speculative design---\emph{structured provocation}, \emph{familiar extrapolation}, \emph{aestheticized noise}, and \emph{buried treasure}---that gives designers and reviewers a shared evaluative vocabulary.
\item A \textbf{practical audit tool}: a reflective checklist for designers to diagnose, evaluate, and improve their speculations (Appendix~A).
\end{enumerate}


\section{Background and related work}

\subsection{Speculation as knowledge generation}
Speculative design is a mode of inquiry that uses designed artifacts to explore possibilities and provoke reflection, emphasizing problem-finding over problem-solving \cite{dunneRaby2013speculativeEverything,auger2013craftingSpeculation,cordova2025slrSpeculative}. Its adjacent traditions---critical design, adversarial design, design fiction---use representational and material practices to interrogate sociotechnical assumptions rather than optimize adoption \cite{dunne2005hertzianTales,disalvo2012adversarialDesign,tanenbaum2012designFictions}. In HCI, this work is articulated as research through design (RtD): knowledge is produced \emph{through} designing and reflecting on artifacts \cite{zimmerman2007rtD,gaver2012expectRtD,schon1983reflectivePractitioner}.

A core tension shapes how we understand speculative work. Pierce distinguishes \emph{progressional} design---``arrow-like,'' oriented toward production and adoption---from \emph{frictional} design, which is ``in tension with progression,'' creating interpretive resistance rather than smooth movement toward implementation \cite{pierce2021tensionProgression}. Frictional work carries \emph{teleological ambiguity}: the artifact's purpose is not to become a product but to enable inquiry and discourse. Pierce identifies five frictional tendencies---\emph{diverging}, \emph{opposing}, \emph{accelerating}, \emph{counterfactualizing}, and \emph{analogizing}---that disrupt progressional trajectories by opening alternatives, resisting dominant trends, pushing to extremes, imagining ``what if'' scenarios, and drawing cross-domain parallels \cite{pierce2021tensionProgression}.

Epistemically, speculative design draws on Cross's ``designerly ways of knowing'' \cite{cross1982designerlyWaysKnowing}, Sch\"{o}n's reflective practice \cite{schon1983reflectivePractitioner}, and philosophical thought experiments. Blythe and Encinas argue that design fictions function analogously to thought experiments, allowing us to test intuitions about unfamiliar situations without building complete systems \cite{blytheEncinas2018researchFiction}. Within RtD, designed artifacts can function as ``strong concepts'' or intermediate-level knowledge: portable abstractions more general than single cases \cite{hookLowgren2012strongConcepts,gaverBowers2012annotatedPortfolios}. Cardenas Cordova et al.'s systematic review identifies four phases---\emph{select}, \emph{explore}, \emph{transform}, \emph{provoke}---suggesting a recognizable process logic across heterogeneous practices \cite{cordova2025slrSpeculative}.

Rahwan, Shariff, and Bonnefon's ``science fiction science'' (sci-fi-sci) method represents a complementary approach from behavioral science: applying controlled experiments to simulated futures---text vignettes, mock applications, virtual environments, physical stagings---to measure how people actually respond to speculated technologies \cite{rahwan2025sciFiSci}. Where qualitative speculative traditions generate rich provocations, sci-fi-sci provides experimental rigor for testing behavioral responses. Epiplexity, we argue, provides the theoretical criteria both traditions need: a principled account of what makes a speculative scenario worth engaging with---qualitatively \emph{or} experimentally.

\subsection{Why speculative quality is hard to judge}
Sterling, a foundational figure in design fiction, observed that while the practice has become ``almost standard,'' ``most design fiction is very bad'' \cite{sterling2013designFiction}---raising the question of what distinguishes insightful speculation from superficial provocation. Critics note that many speculative designs remain ``gallery pieces'' without clear impact \cite{tonkinwise2014reviewDunneRaby}, and that without careful grounding, speculation can reinforce rather than challenge present norms \cite{pradoOliveira2015futuristicGizmos}. HCI has cautioned against judging all design research by usability logics; such criteria can be inappropriate when the goal is exploration or critique \cite{greenbergBuxton2008evaluationHarmful,gaver2012expectRtD}.

Ringfort-Felner et al.'s scoping review crystallizes the problem: quality criteria are distributed across traditions and often implicit \cite{ringfortfelner2025quality}. They propose a taxonomy of nine qualities across three categories: \emph{speculative qualities} (fictional, critical, socio-political), \emph{discursive qualities} (experienceable, thought-provoking), and \emph{process qualities} (grounded, participative, reflected, playful). Yet a persistent question remains: \emph{what are we trying to maximize?} These qualities may conflict---highly fictional work may sacrifice grounding; deeply participative processes can produce diffuse insights under limited time; highly critical work may sacrifice experienceability for polemic. The field needs a way to relate these qualities to a more basic account of \emph{learning} from speculation.

Lindley and Green propose a provocative criterion: ``the ultimate measure of success for speculative design is to disappear completely''---to have its insights absorbed into mainstream thinking \cite{lindleyGreen2022disappear}. This suggests the value lies in \emph{transferable content}: ideas that can be internalized beyond the original artifact. But what determines whether insights transfer? What makes one speculation's lessons ``sticky'' while another's fade? This points toward the need for an account of structured, learnable information in speculative design.

\subsection{Existing approaches to evaluating speculative design}

Despite the acknowledged difficulty, a range of evaluative approaches has emerged. We review them here to clarify what each contributes and what remains unaddressed. The approaches fall into four broad families: taxonomic, typological, analytical, and experimental.

\paragraph{Taxonomic approaches: cataloguing qualities.}
The most systematic effort is Ringfort-Felner et al.'s scoping review, which synthesizes nine qualities across three categories from 63 publications \cite{ringfortfelner2025quality}. Cardenas Cordova et al.'s systematic review contributes a complementary process-level taxonomy---\emph{select}, \emph{explore}, \emph{transform}, \emph{provoke}---but addresses what designers \emph{do} rather than what makes their outputs \emph{good} \cite{cordova2025slrSpeculative}. Baumer, Blythe, and Tanenbaum argue that design fiction is too heterogeneous for unified criteria and instead propose matching evaluation methods to contribution types---critical readings, narratological analysis, studio critique, user studies, or thought experiments \cite{baumer2020evaluatingDesignFiction}. These taxonomic works map the landscape with increasing precision, but they share a common limitation: they describe \emph{what} the field values without explaining \emph{why} certain qualities matter more than others or how to resolve conflicts between them. A speculation cannot simultaneously maximize all nine of Ringfort-Felner et al.'s qualities; the field needs a principle for adjudicating trade-offs.

\paragraph{Typological approaches: classifying design orientations.}
Pierce's progressional/frictional distinction provides the foundational typology: five frictional tendencies---diverging, opposing, accelerating, counterfactualizing, and analogizing---describe \emph{how} speculative work resists progressional vectors \cite{pierce2021tensionProgression}. This is powerful for classification but stops short of evaluation: it tells us what frictional designs \emph{do} but not how well they do it. A speculation can be clearly frictional and still fail to teach anyone anything. Bardzell and Bardzell draw on 150 years of critical theory to develop ``close readings'' of critical designs \cite{bardzell2013criticalDesign}, showing how analytical rigour can be brought to bear---but the resulting evaluations remain interpretive and case-specific rather than generalizable across projects.

\paragraph{Practice-based approaches: design heuristics.}
Auger's perceptual bridge identifies a crucial design variable: speculation must be grounded enough to engage audiences but strange enough to provoke \cite{auger2013craftingSpeculation}. This offers practical guidance for \emph{creating} effective speculations---six bridging techniques for managing the gap between the familiar and the fictional. Yet the bridge metaphor addresses only engagement (can audiences relate?), not learning (what do they extract?). A well-bridged speculation may provoke a strong reaction while leaving observers with nothing transferable. Sterling's influential but informal criterion---design fiction should ``suspend disbelief about change'' through diegetic prototypes \cite{sterling2013designFiction}---operates similarly: it recognizes quality when present but cannot specify the mechanisms that produce it. Tonkinwise's sustained critique provides sharper evaluative teeth, identifying failures of political engagement, diversity, and actionability \cite{tonkinwise2014reviewDunneRaby}, but the critique is destructive rather than constructive: it identifies what is wrong without proposing a principled alternative account of what ``right'' would look like.

\paragraph{Experimental approaches: testing speculative scenarios.}
Rahwan, Shariff, and Bonnefon's science fiction science method represents a fundamentally different strategy: rather than evaluating the \emph{artifact}, it tests \emph{audience responses} through controlled experiments \cite{rahwan2025sciFiSci}. This provides rigorous evidence about how people actually respond to speculated futures across a fidelity spectrum from text vignettes to physical stagings. However, the method does not provide criteria for what makes a speculation \emph{worth testing} in the first place---it presumes that the scenario has already been judged interesting enough to warrant experimental investment.

\paragraph{The persistent gap.}
Table~\ref{tab:framework_comparison} summarises the coverage of existing approaches across four evaluative dimensions. Three systematic gaps emerge:

\begin{enumerate}
\item \textbf{No shared account of what to optimize.} Taxonomies enumerate qualities but provide no principle for choosing among them when they conflict. The field can describe ``good speculation'' in multiple ways but cannot say what makes one description more fundamental than another.
\item \textbf{No account of observer constraints.} With the partial exception of Auger's perceptual bridge and Rahwan et al.'s fidelity spectrum, existing frameworks treat quality as a property of the \emph{artifact} rather than a relation between artifact and \emph{observer}. Yet what people learn depends on who they are, how much time they have, and what scaffolding supports their engagement.
\item \textbf{No distinction between productive and unproductive surprise.} Gaver et al.\ established ambiguity as a design resource over two decades ago \cite{gaver2003ambiguity}, but the critical follow-up question---\emph{when is ambiguity productive and when is it merely confusing?}---has remained unanswered. Existing frameworks cannot distinguish a speculation that generates genuine insight from one that generates only affective shock.
\end{enumerate}

\noindent These gaps motivate our information-theoretic reframing. What is needed is not another taxonomy of qualities but a more fundamental account of what observers \emph{learn} from speculative encounters under realistic constraints---and what determines whether that learning is structured enough to transfer. The remainder of this section develops the theoretical foundations for such an account.

\begin{table*}[h]
\centering
\small
\caption{Coverage of existing evaluation approaches across four dimensions. \ding{51} = explicitly addressed; (\ding{51}) = partially addressed; --- = not addressed.}
\label{tab:framework_comparison}
\begin{tabularx}{\columnwidth}{@{}l >{\centering\arraybackslash}p{1.4cm} >{\centering\arraybackslash}p{1.4cm} >{\centering\arraybackslash}p{1.4cm} >{\centering\arraybackslash}p{1.5cm}@{}}
\toprule
\textbf{Framework} & \textbf{Theoretical basis} & \textbf{Observer\newline awareness} & \textbf{Practical\newline tool} & \textbf{Empirical\newline validation} \\
\midrule
Ringfort-Felner et al.\ \cite{ringfortfelner2025quality} & (\ding{51}) & --- & (\ding{51}) & (\ding{51}) \\
Pierce \cite{pierce2021tensionProgression} & \ding{51} & --- & --- & --- \\
Gaver \cite{gaver2003ambiguity} & \ding{51} & --- & (\ding{51}) & --- \\
Cordova et al.\ \cite{cordova2025slrSpeculative} & (\ding{51}) & --- & (\ding{51}) & (\ding{51}) \\
Auger \cite{auger2013craftingSpeculation} & (\ding{51}) & (\ding{51}) & \ding{51} & --- \\
Tonkinwise \cite{tonkinwise2014reviewDunneRaby} & (\ding{51}) & (\ding{51}) & --- & --- \\
Rahwan et al.\ \cite{rahwan2025sciFiSci} & \ding{51} & \ding{51} & \ding{51} & \ding{51} \\
Baumer et al.\ \cite{baumer2020evaluatingDesignFiction} & (\ding{51}) & --- & (\ding{51}) & --- \\
\midrule
\textbf{Epiplexity (this paper)} & \ding{51} & \ding{51} & \ding{51} & --- \\
\bottomrule
\end{tabularx}
\end{table*}

\subsection{Information theory in design: entropy and two design logics}
Shannon entropy quantifies average uncertainty in a message source: higher entropy means more possible states, more ``surprise'' per observation \cite{shannon1948infoTheory}. Gero and Kan applied this to design processes, developing entropy measures for linkographs showing that higher entropy indicates more diversified ideas and opportunity for quality outcomes \cite{geroKan2007entropyLinkograph,geroKan2018entropyCreativity}. Crucially, entropy captures the \emph{distribution of possibilities}, not their quality.

We can sharpen the progressional/frictional distinction using information-theoretic language. Progressional design is fundamentally \emph{negentropic}: each decision eliminates alternatives, converging toward a single specified artifact. Many design process models (including double-diamond variants) make this convergence logic explicit \cite{designcouncil2005doubleDiamond}. Frictional design is deliberately \emph{entropic}: it maintains uncertainty by introducing alternatives, tensions, and counterfactuals that resist closure.

Ng's account of ``preemptive futures'' develops this framing explicitly \cite{ng2022preemptiveFutures}. Speculation, in this view, is tied to \emph{preemption}---anticipatory action to secure options and mitigate losses in a world where prediction and surprise are entangled. Speculative design becomes ``the increase in entropy or the maximization of surprises in a system'' \cite{ng2022preemptiveFutures}. Rather than forecasting a single future, it multiplies possibilities to reveal dependencies, vulnerabilities, and intervention points.

However, \emph{raw entropy is not the same as value}. A random scenario generator produces maximal entropy without insight. We need to distinguish \emph{structured} uncertainty that enables learning from \emph{unstructured} noise. As Frederik Pohl observed---and as Rahwan et al.\ foreground---``a good science fiction story should be able to predict not the automobile but the traffic jam'' \cite{rahwan2025sciFiSci}: what matters is not the technology itself but the second-order social consequences it generates. In the language of this paper, the traffic jam is $S_t$---structured, transferable insight about how systems reshape behavior---while the automobile is surface-level novelty that contributes primarily to $H_t$. This distinction---between productive surprise and mere randomness---is the central concern of the next section.

\subsection{Structured complexity and the cognitive science of productive surprise}

A convergent body of work across complexity science, psychology, and neuroscience establishes that valuable information lies at the boundary between order and randomness. Langton's ``edge of chaos'' identifies a critical zone where complex systems exhibit maximal computational capacity---neither frozen nor chaotic, but poised where structure and surprise coexist \cite{langton1990edgeChaos}. Berlyne's arousal theory established the inverted-U: stimuli of intermediate complexity generate optimal engagement \cite{berlyne1960conflictArousal}. Schmidhuber formalizes ``interestingness'' as the rate of compression progress---only \emph{learnable-but-not-yet-learned} structure sustains engagement \cite{schmidhuber2010formalCreativity}. Itti and Baldi's Bayesian surprise measures how much data changes beliefs, showing that random noise carries high Shannon information but virtually zero belief-updating impact \cite{ittiBaldi2009bayesianSurprise}. Silvia's appraisal theory adds that interest requires stimuli appraised as both novel \emph{and} comprehensible \cite{silvia2005interestAppraisal}. The predictive processing framework provides the neurocognitive architecture: the brain minimizes prediction error, and aesthetic engagement accompanies the resolution of \emph{reducible ambiguity}---prediction errors the observer can eventually resolve \cite{friston2010freeEnergy,vanDeCruys2011predictionArt}. For speculative design, the extension is from perceptual prediction error (surprise at what you \emph{see}) to epistemic prediction error (surprise at what you \emph{think is possible}).

Gaver, Beaver, and Benford's ``Ambiguity as a Resource for Design'' reframed ambiguity from design failure to design strategy \cite{gaver2003ambiguity}. But the critical question has remained unanswered since 2003: \emph{how do you distinguish productive ambiguity from mere confusion?} This is precisely the gap that epiplexity addresses. All frameworks converge on one principle: \textbf{optimal cognitive engagement occurs at intermediate levels of reducible complexity, where bounded observers encounter structure that rewards effort}. This resonates with Auger's ``perceptual bridge'': speculation fails if too fantastical (pure noise) or too familiar (no surprise) \cite{auger2013craftingSpeculation}. Quality in speculative design thus becomes a problem of information allocation: given bounded resources, what aspects of a future-space does an artifact help people reliably infer?

\subsection{Epiplexity: structured information for bounded observers}
Classical information measures abstract away observer constraints. Shannon entropy $H(X)$ quantifies uncertainty assuming idealized channels; Kolmogorov complexity $K(x)$ quantifies description length assuming unbounded computation. But real observers are bounded. Finzi et al.\ introduce \emph{epiplexity} (from epistemic complexity) to formalize what computationally bounded observers can actually learn from data \cite{finzi2026epiplexity}.

Formally, given a random variable $X$ and a computational time bound $t$, epiplexity $S_t(X)$ is defined as the description length of the optimal probabilistic model that minimizes total description length under computational constraints, grounded in Minimum Description Length (MDL) theory augmented with time bounds drawn from cryptography \cite{finzi2026epiplexity}. The total description length decomposes into:
$$L_t(X) = S_t(X) + H_t(X)$$
where $S_t(X)$ is the \emph{epiplexity}---the structured, learnable patterns that a time-bounded observer can extract---and $H_t(X)$ is the \emph{time-bounded entropy}---residual randomness that appears as noise to any observer operating within budget $t$. As $t \to \infty$, $S_t$ approaches the total learnable structure; for finite $t$, much potential structure remains inaccessible.

\begin{quote}
\textbf{Epiplexity (Definition):} From \emph{epi-} (upon) + \emph{-plexity} (complexity/perplexity): ``epistemic complexity.'' The minimum description length of data achievable within a time-bounded computation, capturing the structural information---learnable patterns, regularities, and order---that a computationally bounded observer can extract. Unlike Shannon entropy (which measures total uncertainty assuming ideal channels) or Kolmogorov complexity (which assumes unbounded computation), epiplexity is inherently observer-relative: the same data can have high epiplexity for one computational budget and low epiplexity for another.
\end{quote}

The concept resolves three paradoxes in classical information theory: for bounded observers, deterministic transformations \emph{can} create information, data ordering \emph{does} matter, and models \emph{can} develop representations richer than the generating process \cite{finzi2026epiplexity}. The intellectual lineage runs through Simon's bounded rationality \cite{simon1955boundedRationality}, Kahneman and Tversky's work on cognitive constraints \cite{kahnemanTversky1979prospect}, and Pirolli and Card's information foraging theory \cite{pirolliCard1999infoForaging}. This tradition establishes that \emph{learning is always constrained}: what people extract depends not only on the source but on the observer's computational and attentional budget.

We propose translating epiplexity---\emph{by analogy}---into speculative design. Speculative artifacts are ``data'' for human sensemaking, but observers are bounded in ways that shape what they can learn. The key question is not whether a scenario contains ``a lot of information'' absolutely, but whether it enables observers to extract \emph{structured} insight that transfers across plausible futures given realistic constraints. A high-epiplexity speculation yields patterns---about incentives, governance, values, invariants---that observers can identify and reuse; a low-epiplexity speculation may be surprising but leaves observers with little they can articulate or apply elsewhere.

This translation is \emph{perspectival, not formal}: we do not claim to quantify epiplexity in speculative design with mathematical precision. The framework provides a \emph{lens}---questions to ask about what observers can learn from a speculation, given their constraints.


\section{Modeling speculative design through information theory}

The central question this paper addresses is: \textbf{What can people actually learn from your speculation?} Not just ``what will they feel'' or ``will they be surprised''---but what structured, reusable insights will they walk away with? This section develops the theoretical model; the next section provides practical tools for applying it.

\subsection{Speculative design as a bounded information process}

We model speculative design as a \emph{bounded information process}. A speculative artifact---scenario, prop, enactment, prototype---encodes a partial model of a possible future. An observer---participant, reader, reviewer, policymaker---engages with that artifact under constraints: limited time, limited background knowledge, limited attention. The observer's task is to extract useful understanding from the encounter.

\begin{figure*}[h]
\centering
\includegraphics[width=0.9\linewidth]{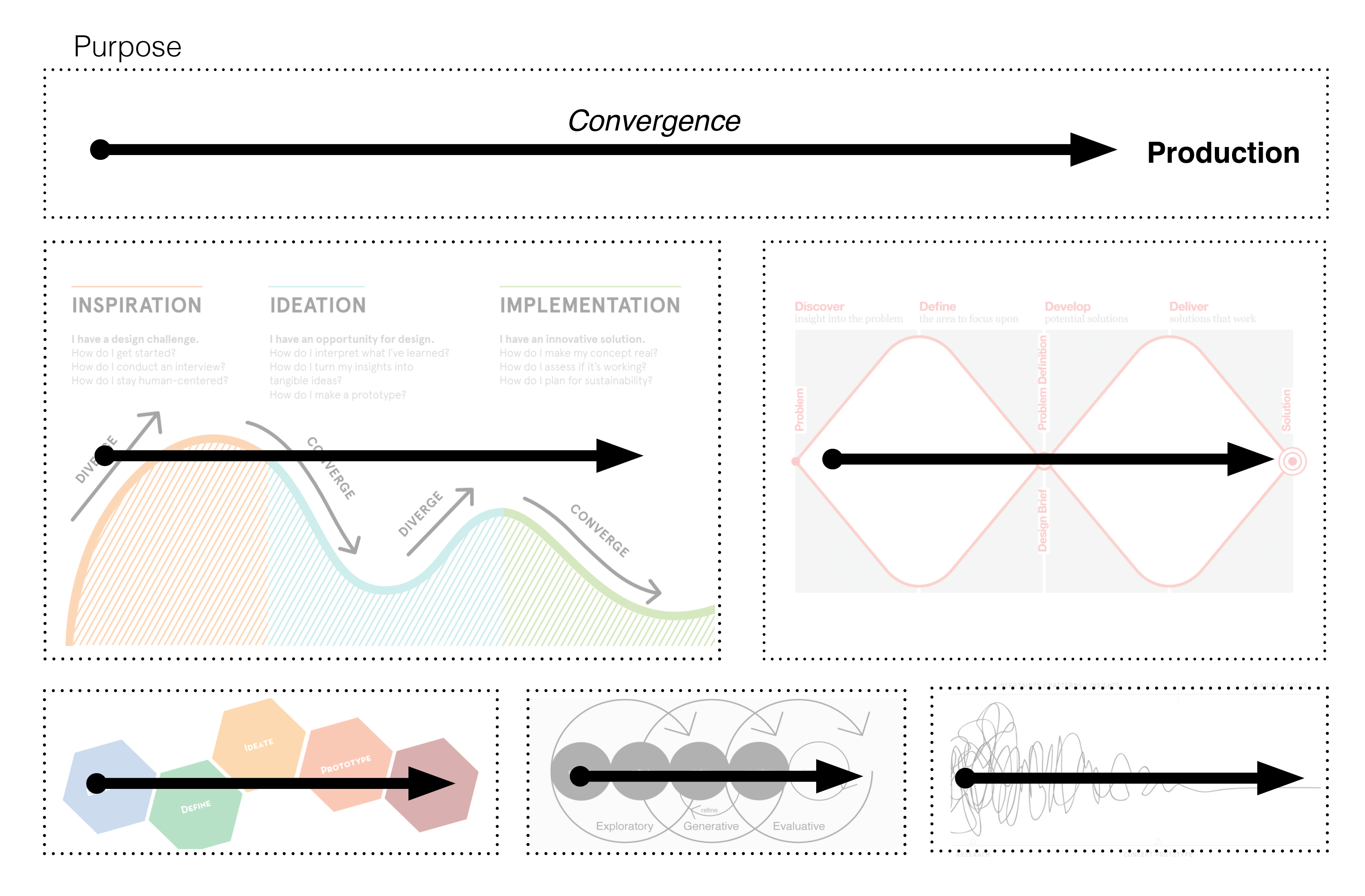}
\Description{A composite diagram showing six progressional design process models, all sharing a common horizontal arrow labeled ``Convergence'' pointing from ``Purpose'' on the left to ``Production'' on the right. The six panels illustrate variants of the design process (including IDEO's three-phase inspiration--ideation--implementation model, the British Design Council's double diamond, and others), each depicting how design activity narrows from divergent exploration to convergent production over time. All models share the same underlying trajectory: a left-to-right movement from open-ended possibility toward a single realized artifact.}
\caption{Progressional design converges toward production through successive phases of inspiration, ideation, and implementation. Reproduced from Pierce \cite{pierce2021tensionProgression}, Figure 2f.}
\label{fig:pierce_progressional}
\end{figure*}

\begin{figure*}[h]
\centering
\includegraphics[width=0.9\linewidth]{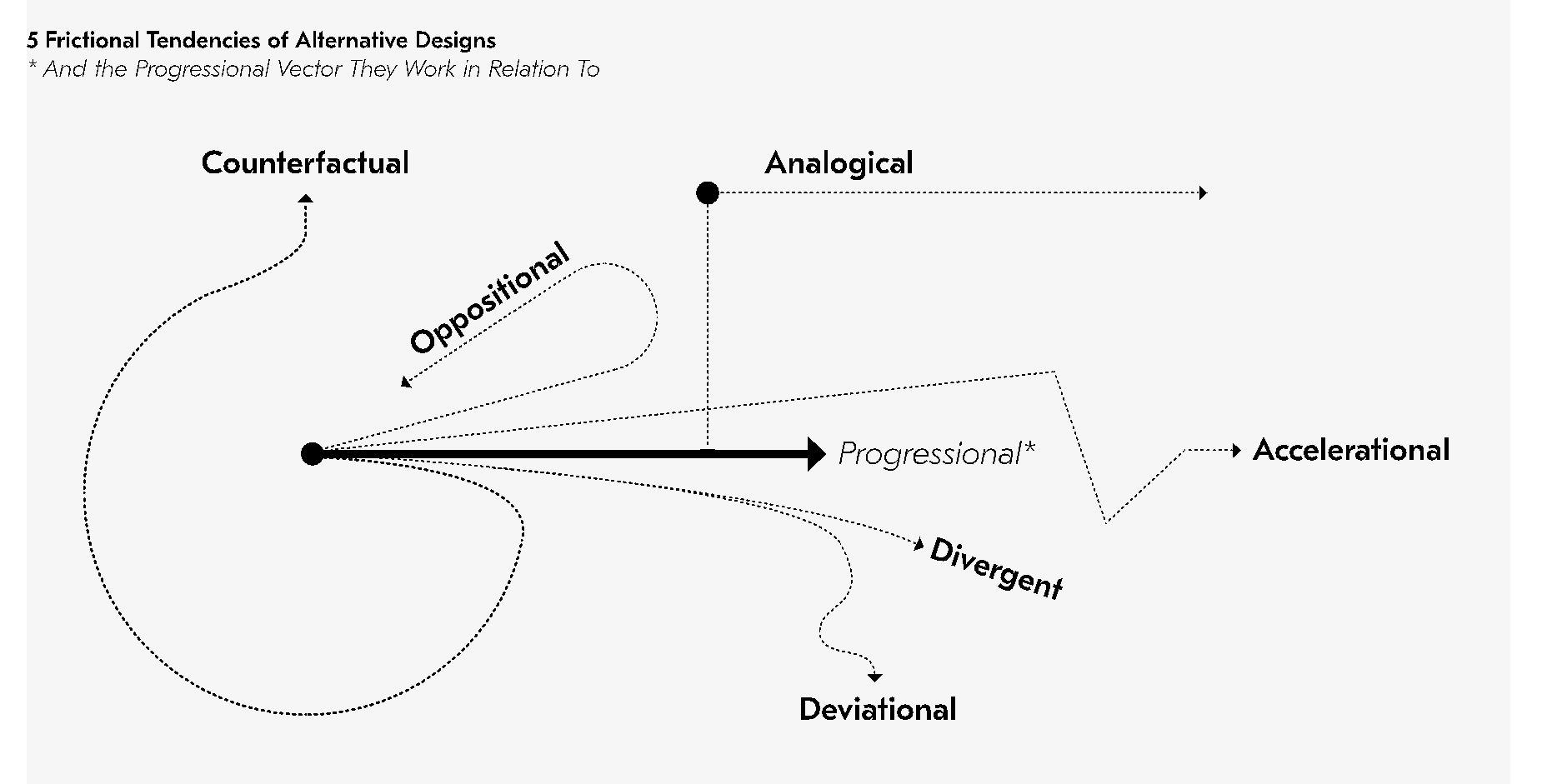}
\Description{A spatial diagram showing a bold horizontal arrow labeled ``Progressional'' pointing to the right, representing the dominant convergent design trajectory. Five frictional tendencies radiate away from this arrow in different directions, each depicted as a dashed curved or angled line: ``Counterfactual'' loops backward and upward to the left, imagining alternative pasts; ``Analogical'' extends vertically upward from a point on the progressional axis, drawing cross-domain parallels; ``Oppositional'' curves back against the progressional direction, resisting dominant trends; ``Divergent'' angles downward and to the right, opening alternative paths; and ``Deviational'' drops downward, departing from the main trajectory. ``Accelerational'' extends further to the right beyond the progressional arrow, pushing trends to their extreme. Together, the five frictional tendencies illustrate how alternative design practices work in tension with the progressional vector rather than along it.}
\caption{Five frictional tendencies of alternative designs---counterfactual, analogical, oppositional, divergent, deviational, and accelerational---and the progressional vector they work in relation to. Reproduced from Pierce \cite{pierce2021tensionProgression}.}
\label{fig:pierce_frictional}
\end{figure*}

\begin{figure*}[h]
\centering
\includegraphics[width=0.9\linewidth]{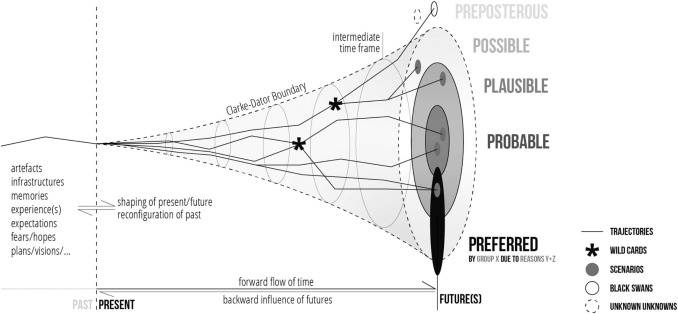}
\Description{A cone-shaped diagram expanding from left to right along a horizontal time axis running from ``Past'' and ``Present'' on the left to ``Future(s)'' on the right. The cone widens into nested concentric zones of increasing uncertainty: a narrow dark core labeled ``Probable,'' surrounded by a lighter zone labeled ``Plausible,'' then ``Possible,'' and an outermost zone labeled ``Preposterous.'' A wedge labeled ``Preferred'' is highlighted within the probable zone, annotated ``by group X, due to reasons Y--Z.'' The left side of the diagram lists elements that shape present understanding: artefacts, infrastructures, memories, experiences, expectations, fears, hopes, plans, and visions. A dashed ``Clarke--Dator Boundary'' line separates the plausible from the possible zone. Asterisk symbols mark ``wild cards'' (low-probability, high-impact events), circles mark ``black swans,'' and a diamond marks ``unknown unknowns.'' Dotted lines represent individual future trajectories fanning outward through the cone. An ``intermediate time frame'' annotation marks a cross-section of the cone. Arrows along the bottom indicate both the forward flow of time and the backward influence of futures on present decisions.}
\caption{The revised futures cone showing how speculative design opens possibility space across probable, plausible, possible, and preposterous futures. Reproduced from Gall et al.\ \cite{gallValletYannou2022futuresCone}, Figure 6.}
\label{fig:futures_cone_revised}
\end{figure*}

This model requires understanding two fundamentally different design logics. Pierce distinguishes \emph{progressional} design---convergent, ``arrow-like,'' oriented toward production (Figure~\ref{fig:pierce_progressional})---from \emph{frictional} design that resists that vector (Figure~\ref{fig:pierce_frictional}) \cite{pierce2021tensionProgression}. Progressional design is negentropic: each decision eliminates alternatives, converging toward a single artifact \cite{designcouncil2005doubleDiamond}. Speculative design works differently---it deliberately keeps possibilities open. The futures cone (Figure~\ref{fig:futures_cone_revised}) visualizes this: trajectories fan outward into possible, plausible, and preferable futures \cite{voros2003foresightProcess,gallValletYannou2022futuresCone}. The goal is not to predict or build, but to \emph{surface what is at stake}---to increase the legibility of a future-space so that observers can reason about trajectories, risks, and interventions \cite{ng2022preemptiveFutures}.

This framing matters because frictional design resists the evaluation criteria that work for progressional design. We cannot ask ``does it reduce uncertainty toward a product?'' because frictional work deliberately maintains uncertainty. But this does not mean frictional work is unevaluable---it means we need different criteria. Epiplexity provides such criteria: it asks not whether uncertainty is reduced toward implementation, but whether the maintained uncertainty yields \emph{learnable structure} about possibility space \cite{pierce2021tensionProgression,finzi2026epiplexity}. A speculation can be deliberately open-ended and still be well-designed for learning. The analogy to computational learning is productive: just as a machine learning model extracts patterns from a dataset within a compute budget, a human observer extracts patterns from a speculative artifact within a cognitive budget. The value of the encounter depends not on the total ``information'' present, but on what the bounded agent can actually learn. Rahwan et al.'s analysis of validity threats in science fiction science provides empirical evidence for this bounded-observer model: participants struggle to simulate unfamiliar decision contexts (the \emph{participant simulation gap}), depicted technologies inevitably diverge from actual implementations, and social contexts shift between the time of speculation and the time of realization \cite{rahwan2025sciFiSci}. These are not merely methodological inconveniences but empirical manifestations of the observer-boundedness that epiplexity formalizes: what people can extract from a speculative encounter is constrained by their cognitive and experiential budget $t$.

\subsection{The epiplexity decomposition in speculative design}

Opening up possibilities does not, by itself, guarantee learning. A random scenario generator produces endless surprises but teaches nothing; shock without structure is noise. The cognitive science reviewed in Section 2.4 converges on this point: random stimuli carry high Shannon information but virtually zero Bayesian surprise \cite{ittiBaldi2009bayesianSurprise}, enable no compression progress \cite{schmidhuber2010formalCreativity}, and produce vicious rather than virtuous confusion cycles \cite{dmelloGraesser2014confusion}. What distinguishes a speculation that leaves people saying ``that's creepy'' from one that equips them with transferable insight about governance, incentives, or contestability is not elaboration but \emph{structure}---patterns that persist across variations of the scenario and compress into reusable understanding.

Epiplexity provides the formal anchor for this distinction. Recall from Section 2.5 that epiplexity decomposes information into structured patterns ($S_t$) that bounded observers can extract and residual entropy ($H_t$) that remains noise given their constraints. Applied to speculative design:

\begin{itemize}
\item $S_t$ (\emph{structured information}): Learnable patterns the observer can extract and generalize---second-order effects, value tensions, governance questions, boundary conditions. These are insights people can articulate and apply elsewhere.
\item $H_t$ (\emph{residual entropy}): Contingent detail, aesthetic noise, complexity that exceeds the observer's extractive capacity. This may include striking imagery, elaborate worldbuilding texture, or scenario specifics that do not compress into reusable insight.
\end{itemize}

A high-epiplexity speculation yields patterns people can articulate and apply elsewhere. A low-epiplexity speculation provokes reactions but leaves people with nothing transferable. This framing shifts the evaluation question from ``Is it surprising?'' to ``Is it \emph{learnably} surprising?''

The decomposition yields a natural diagnostic space defined by two axes---$S_t$ and $H_t$---that distinguishes four qualitatively different outcomes of speculative design:

\begin{enumerate}
\item \textbf{High epiplexity, calibrated entropy.} The ideal. Rich extractable structure that rewards engagement, with genuine uncertainty that compels interpretation. The artifact operates at Auger's ``perceptual bridge'' \cite{auger2013craftingSpeculation}---plausible enough to engage but strange enough to provoke. Observers encounter learnable-but-not-yet-learned structure, sustaining what Schmidhuber formalizes as compression progress \cite{schmidhuber2010formalCreativity}.

\item \textbf{Low epiplexity, low entropy.} Obvious, shallow speculation. Futures that merely extrapolate present trends without structural complexity---the ``smart fridge'' problem. This is what Tonkinwise critiques as futures operating within a ``shopping framework'' \cite{tonkinwise2014reviewDunneRaby}: both $S_t$ and $H_t$ are low because there is little to learn and little surprise.

\item \textbf{Low epiplexity, high entropy.} Aestheticized noise. Visually or conceptually complex surfaces with no extractable structure underneath. The design equivalent of a cryptographically secure pseudorandom generator: it looks complex but contains trivial information for any bounded observer \cite{finzi2026epiplexity}. An artifact that is merely stylistically provocative---dramatic aesthetics, shocking imagery---without underlying extractable structure registers as high $H_t$ but low $S_t$. This provides a non-subjective criterion for the critique that first-wave speculative design often produced ``fashion editorial'' provocations \cite{tonkinwise2014reviewDunneRaby,sterling2013designFiction}: the question shifts from ``is this shocking?'' to ``is there learnable structure here that rewards cognitive investment?''

\item \textbf{High epiplexity, excessive entropy.} Dense but chaotic. Potentially rich structure that is practically inaccessible because the noise floor overwhelms the signal. Genuine insight is buried in incoherence---the $S_t$ is present but the engagement budget $t$ is insufficient for extraction. This is not necessarily a design failure but a curatorial or facilitation challenge: the same artifact may shift from this quadrant to the first when scaffolding increases the observer's effective budget \cite{dmelloGraesser2014confusion,bjork1994desirableDifficulties}.
\end{enumerate}

These quadrants are developed further in Section 4 with illustrative examples.

The decomposition also suggests a design analog of Finzi et al.'s practical estimation methods. In machine learning, the ``prequential coding'' heuristic approximates epiplexity as the area between a model's initial training loss and its final asymptotic loss---intuitively, the cumulative learning that occurs as a model extracts structure from data \cite{finzi2026epiplexity}. The design analog would track how audience understanding evolves during sustained engagement with an artifact. A high-epiplexity artifact would show continuous learning: diminishing but persistent insight extraction over time, as observers progressively compress the speculation into reusable patterns. A low-epiplexity artifact would show either immediate comprehension (no learning curve---familiar extrapolation) or a flat line (no learning at all, just persistent confusion---aestheticized noise). While we do not propose formal measurement here, this suggests concrete empirical operationalizations: think-aloud protocols tracking interpretive progression, longitudinal studies of how extracted insights evolve across multiple encounters, or analysis of whether audience responses converge on structural themes over time.

\subsection{What counts as structured information in speculative design}

The epiplexity decomposition raises a concrete question: what \emph{kinds} of structure can bounded observers actually extract from speculative design artifacts? In machine learning, structure means compressible regularities---patterns a model can exploit to reduce prediction error. In speculative design, the relevant structures are epistemic: they concern how sociotechnical systems work, who they affect, and what conditions shape outcomes across different futures. We identify four recurring forms of structured information ($S_t$) that are particularly valuable because they compress into transferable insight:

\begin{enumerate}
\item \textbf{Second-order effects and causal coupling.} How do systems reshape behavior, labor, and power through incentives and institutional feedback loops? A speculation that makes incentive structures \emph{inferable}---not merely asserted---lets observers trace causal chains from technological interventions to social consequences \cite{rosenblatStark2016uberBigDataSociety,graySuri2019ghostWork,eubanks2018automatingInequality}. A high-$S_t$ speculation encodes these coupling mechanisms so that observers can extract them from the scenario itself, rather than needing them to be didactically explained.

\item \textbf{Value tensions and moral framings.} Competing values that are otherwise latent---privacy vs.\ convenience, autonomy vs.\ safety, efficiency vs.\ justice---become extractable structure when a speculation forces observers to confront trade-offs rather than resolve them prematurely \cite{friedmanHendry2019VSD,costanzaChock2020designJustice}. The value tension is structural precisely because it persists: changing the specific technology does not dissolve the dilemma.

\item \textbf{Governance and contestability conditions.} Who can dispute, audit, or refuse? What institutional arrangements enable or foreclose accountability? These questions constitute learnable structure because they identify \emph{invariant governance requirements}---conditions that any legitimate deployment of a technology must address regardless of implementation details \cite{alfrink2023contestableCameraCars,mitchell2019modelCards}.

\item \textbf{Boundary conditions and persistent constraints.} Data locality, information asymmetries, infrastructural path dependence, and asymmetric power over technical standards constitute constraints that persist across different futures \cite{mortier2014humanDataInteraction,zuboff2019surveillanceCapitalism}. These are the structural ``walls'' of possibility space---they compress well because they hold across many scenarios.
\end{enumerate}

These four forms share a common property: they are \emph{robust to superficial variation}. Change the interface, the brand, the geography, or the aesthetic wrapper, and the structure persists. This robustness is precisely what makes them compressible in the MDL sense---they are regularities that a bounded observer can identify and reuse across contexts, rather than contingent details that apply only to one scenario.

The connection to epistemic objects clarifies why structure matters. Boserman argues, drawing on Rheinberger's concept of \emph{epistemic things}, that speculative design prototypes function as objects characterized by productive incompleteness---they embody ``what one does not yet know'' and generate knowledge through experimental engagement \cite{boserman2019epistemicObjects}. The critical distinction is between \emph{structured incompleteness} and \emph{random indeterminacy}: an epistemic object is generative precisely because its gaps are patterned, inviting specific forms of inquiry rather than diffuse confusion. In epiplexity terms, a well-designed speculative artifact has high $S_t$ because its incompleteness is structured---the missing pieces constrain what observers can hypothesize, directing interpretive effort toward productive territory. By contrast, an artifact whose incompleteness is random (high $H_t$, low $S_t$) offers no such direction: observers may generate hypotheses, but the artifact provides no foothold for selecting among them. This distinction also sharpens the boundary between \emph{epistemic objects} and \emph{boundary objects} \cite{starGriesemer1989boundaryObjects}: boundary objects coordinate across communities through stability (low entropy), while epistemic objects generate new knowledge through structured openness (high epiplexity).

This connects to the logic of \emph{abductive reasoning}---inference from surprising observations to explanatory hypotheses---which is the fundamental cognitive mode that speculative design triggers \cite{kolko2010abductiveThinking,dorst2011designThinking}. The quality of abduction a speculation supports depends on the relationship between its structured information and its residual noise. Too much structure with too little entropy (Quadrant~II) constrains abduction to a single predetermined interpretation---the observer is told what to conclude. Too much noise with too little structure (Quadrant~III) provides no foothold for abduction at all---the observer encounters surprise but cannot generate testable hypotheses about why things are as they are. The productive zone (Quadrant~I) enables genuine hypothesis generation: enough structure that observers can reason abductively about causal mechanisms, governance failures, and value conflicts, with enough entropy that the reasoning remains open and contestable.

A high-epiplexity speculation thus makes structures \emph{inferable}---not by explaining them didactically, but by designing provocations that force interpretation at structural levels . This is the speculative design analog of defamiliarization: the familiar is made strange in a way that reveals structure, rather than in a way that merely disorients \cite{shklovsky1917artTechnique,bellBlytheSengers2005defamiliarization}. Suvin's ``cognitive estrangement'' captures the same principle from science fiction theory: the \emph{novum} (radically new element) must be simultaneously estranging \emph{and} cognitive---it disrupts existing models while providing sufficient regularity for new models to form \cite{suvin1979metamorphoses}. In the language of this paper, cognitive estrangement \emph{is} high epiplexity: structured surprise that a bounded observer can compress into reusable understanding.

\subsection{Observer-dependence: for whom, with what resources, what transfers}

A crucial feature of epiplexity---and a central contribution of this paper---is its explicit \emph{observer-dependence}. Unlike quality criteria that aspire to universal validity, epiplexity acknowledges that what people learn from a speculation depends on who they are, what resources they bring, and what they do with the insight afterward. This is not a limitation but a design consideration to be embraced.

The epiplexity framing translates into three practical questions:

\paragraph{1. For whom?} Who is your intended audience? A designer learning about problem spaces extracts different patterns than a policymaker anticipating governance needs, who learns differently than a public reflecting on values. Clarify your target: Who should learn from this? What do they already know?

\paragraph{2. With what resources?} What is the realistic engagement budget? A speculation that works in a facilitated workshop may fail in a portfolio scroll. Consider: How much time will people spend? What scaffolding (facilitation, documentation, props) supports interpretation?

\paragraph{3. What transfers?} Will the insight matter next year? Does it apply beyond this specific scenario? The most valuable patterns are \emph{robust} (persist if superficial details change) and \emph{reusable} (can be articulated as design principles, policy questions, or governance considerations that transfer) \cite{hookLowgren2012strongConcepts}.

This observer-dependence means an artifact's epiplexity is not fixed: the same work may be richly informative in a facilitated workshop and opaque in a gallery walkthrough. As D'Mello and Graesser's work shows, whether confusion becomes productive or destructive depends on scaffolding and the observer's capacity for resolution \cite{dmelloGraesser2014confusion}. Rahwan et al.'s fidelity spectrum of simulation methods---from text vignettes through mock applications and virtual reality to physical stagings---operationalizes this same principle \cite{rahwan2025sciFiSci}. Each step up the fidelity ladder expands the observer's effective engagement budget $t$: a text vignette provides minimal perceptual scaffolding, risking Quadrant~IV outcomes where structure remains buried; an immersive VR environment or physical staging extends the budget, enabling extraction of richer causal and institutional structure. The choice of simulation fidelity is, in epiplexity terms, a choice about how much computational resource to allocate to the observer. This is not a flaw in the framework but an actionable insight: \emph{design the engagement, not just the artifact}.


\section{Auditing speculative design for epiplexity}

Section 3 developed a theoretical model. This section provides practical tools: a diagnostic map for understanding where a speculation falls, and a reflective audit for evaluating and improving speculations. We use illustrative examples throughout to ground the framework.

\paragraph{Methodological note.} The examples in this section are \emph{illustrative}, not confirmatory. We selected cases retrospectively to demonstrate how the framework applies, not to test whether it predicts outcomes. Cases were chosen for diversity of domain and because they are well-documented in prior literature. Our assessments reflect our interpretation as informed readers; systematic empirical validation remains future work.

\subsection{A diagnostic map: four quadrants of speculative design quality}

\begin{figure*}[h]
\centering
\includegraphics[width=0.8\linewidth]{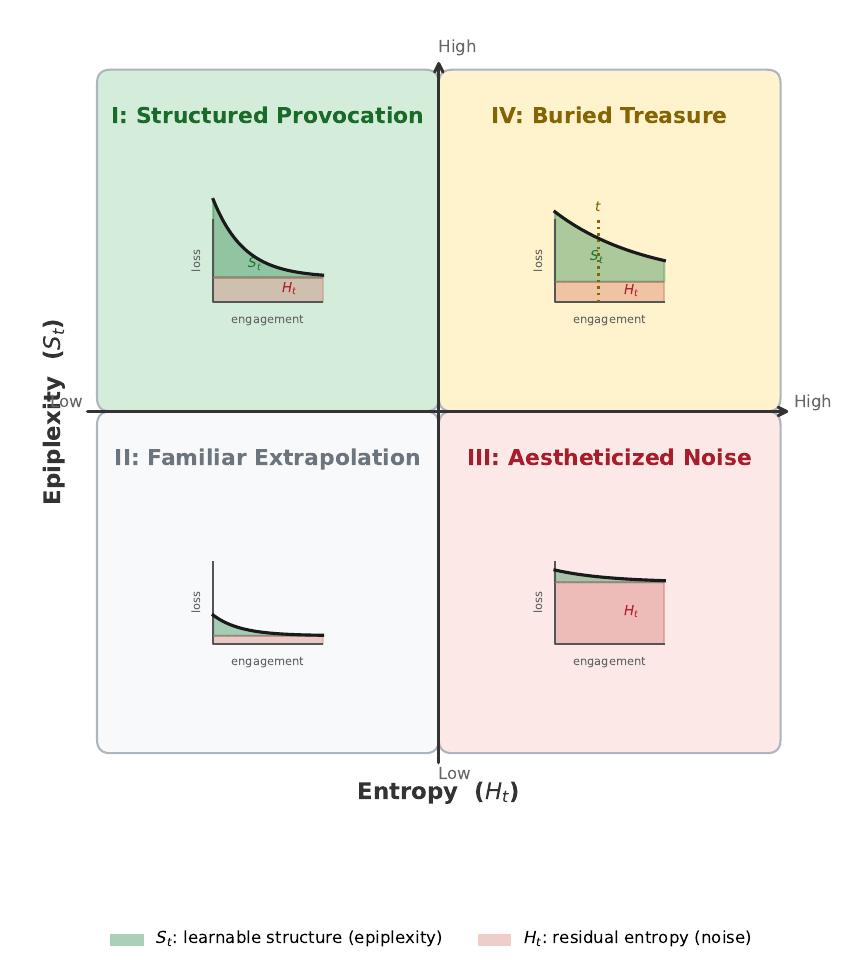}
\Description{A two-by-two matrix with Epiplexity (S\_t) on the vertical axis (low to high, bottom to top) and Entropy (H\_t) on the horizontal axis (low to high, left to right), defining four color-coded quadrants. Each quadrant contains a small inset chart plotting a learning curve (loss versus engagement) that visualizes the decomposition into learnable structure (S\_t, shaded green) and residual noise (H\_t, shaded pink). Quadrant I (top-left, green): ``Structured Provocation''---the learning curve drops steeply, showing a large green S\_t area above a modest pink H\_t band, indicating rich extractable structure with calibrated surprise. Quadrant IV (top-right, yellow): ``Buried Treasure''---a similar curve shape but a vertical dashed line labeled t truncates extraction early, indicating that genuine structure exists but the observer's time budget is exhausted before it can be fully extracted; H\_t remains large. Quadrant II (bottom-left, white): ``Familiar Extrapolation''---the curve is nearly flat at a low level, with thin S\_t and H\_t bands, indicating little surprise and little to learn. Quadrant III (bottom-right, pink): ``Aestheticized Noise''---the curve stays flat and high, with a large pink H\_t region and negligible green S\_t, indicating high surprise but no learnable structure. A legend at the bottom identifies the green fill as ``S\_t: learnable structure (epiplexity)'' and the pink fill as ``H\_t: residual entropy (noise).''}
\caption{Four quadrants of speculative design quality. The vertical axis represents epiplexity ($S_t$)---how much structured, learnable information bounded observers can extract. The horizontal axis represents entropy ($H_t$)---the degree of surprise or noise. Effective speculation occupies Quadrant~I: high epiplexity with calibrated entropy.}
\label{fig:quadrants}
\end{figure*}

The epiplexity model yields a natural diagnostic map (Figure~\ref{fig:quadrants}). Two axes---the degree of surprise or entropy an artifact generates (how unfamiliar and disorienting the encounter is) and the degree of epiplexity (how much of that surprise contains learnable, transferable structure)---define four qualitatively distinct outcomes. These quadrants are not categories to impose but a \emph{diagnostic vocabulary} for reflection.

\subsubsection{Quadrant I: Structured Provocation (High Epiplexity, Calibrated Entropy)}
Rich extractable structure that rewards engagement, with genuine uncertainty that compels interpretation. The artifact operates at the ``edge of chaos'' \cite{langton1990edgeChaos}---strange enough that observers must build new models, structured enough that they can. This is D'Mello and Graesser's ``virtuous cycle'': productive confusion that resolves into deeper understanding \cite{dmelloGraesser2014confusion}.

\emph{Feels like:} ``I didn't expect that---but now I see why it would happen, and it changes how I think about this.''

\emph{Illustrative examples.} ``Contestable Camera Cars'' frames public AI as open to dispute, embedding governance tensions---contestability, legitimacy, evidence standards---that persist even if the specific sensing modality changes \cite{alfrink2023contestableCameraCars}. Swap the camera car for drones or wearable sensors: the structural questions remain. Superflux's ``Mitigation of Shock'' constructs a future London apartment adapted to climate disruption, grounding speculation in material constraints (modified furniture, preserved foods, off-grid energy) that make adaptation strategies inferable and policy-relevant \cite{superflux2017mitigationShock}. The Near Future Laboratory's ``TBD Catalog'' embeds speculative technologies within mundane consumer contexts, making visible second-order effects and behavioral adaptations that transfer beyond any specific product \cite{nearfuture2014tbdCatalog}. In each case, $S_t$ is high because the work foregrounds \emph{forces}---governance gaps, institutional inertia, incentive structures---not merely technologies.

\subsubsection{Quadrant II: Familiar Extrapolation (Low Epiplexity, Low Entropy)}
Incremental speculation that extends present trends without structural complexity. No real surprise, no deep insight. The ``smart fridge'' problem.

\emph{Feels like:} ``Yes, and? This is basically what we have now but shinier.''

\emph{Illustrative examples.} Speculative concepts that add ``AI'' to existing products without interrogating what restructures: a smarter voice assistant, a faster recommendation engine, a more responsive smart home. Such work confirms existing assumptions rather than challenging them---what Tonkinwise critiques as futures operating within the ``shopping framework'' \cite{tonkinwise2014reviewDunneRaby}. Both $S_t$ and $H_t$ are low: observers learn little because there is little to learn, and little surprise because the scenario is already familiar.

\subsubsection{Quadrant III: Aestheticized Noise (Low Epiplexity, High Entropy)}
Visually or conceptually complex surfaces with no extractable structure underneath. Strong affect, zero transfer. Shock is the entire content. The design equivalent of a cryptographically secure pseudorandom generator: it looks complex but contains trivial information for any bounded observer \cite{finzi2026epiplexity}.

\emph{Feels like:} ``That's disturbing---but I can't tell you what I'm supposed to do with that.''

\emph{Illustrative examples.} ``Gadget dystopias'' that scale up a current technology (e.g., facial recognition) into a nightmare future without specifying institutions, incentives, or contestability mechanisms. Observers conclude ``surveillance is bad''---something they already knew---but gain no tools for reasoning about \emph{conditions under which} harms emerge or can be mitigated. The causal model is underdeveloped; $H_t$ is high (contingent scenario details, dramatic aesthetics) while $S_t$ is low. Similarly, a speculative AI companion concept that simply anthropomorphizes an LLM (``a holographic friend who talks like you'') may elicit affect but remain structurally thin: observers focus on novelty or persona without converging on reusable insight about data governance, incentive misalignment, or algorithmic authority over personal narrative. Sterling's observation that ``most design fiction is very bad'' \cite{sterling2013designFiction} targets this quadrant: provocative surfaces that cannot be compressed into reusable understanding.

\subsubsection{Quadrant IV: Buried Treasure (High Epiplexity, Excessive Entropy)}
Genuine structural richness that is practically inaccessible because the noise floor overwhelms the signal. The insight exists but bounded observers cannot extract it within realistic engagement budgets.

\emph{Feels like:} ``I sense there's something important here, but I can't figure out what it is in the time I have.''

\emph{Illustrative examples.} Deliberately uncomfortable artifacts---a provocatively designed object that disrupts norms of comfort and consumption---may contain genuine insight about bodily expectation, consumer habituation, and designed compliance, but under typical gallery engagement (30 seconds) observers converge on the surface message (``comfort is taken for granted'') rather than the deeper causal structures \cite{pierce2021tensionProgression,wakkaryMaestri2007resourcefulness}. The $S_t$ is present but the $t$ (engagement budget) is insufficient. Similarly, dense participatory speculations where rich structural content is generated but diffused across too many threads, with no curatorial frame to help observers compress, may produce Q4 outcomes: meaningful process, inaccessible output. Speculative work in sensitive social domains---such as ``Magic Machines for Refugees'' \cite{almohamed2020magicMachinesRefugees}---can surface rich infrastructural and bureaucratic constraints shaping displacement and agency, but the insight may require facilitation and contextual knowledge that casual observers lack.

Q4 is not necessarily a design failure---it may be a \emph{curatorial} or \emph{facilitation} challenge. The same artifact can shift from Q4 to Q1 when the engagement context changes: a facilitated workshop, documentation layer, or interpretive scaffold may make buried structure extractable. This connects to the growing use of speculative methods in policy contexts, where the challenge is often not generating rich speculation but making its insights accessible to time-constrained policymakers \cite{tsekleves2022speculativePolicy}.

\subsection{The epiplexity audit: a practical tool for designers}

The four quadrants provide a diagnostic map; to make it actionable, we developed a comprehensive reflective checklist organized around the framework's four conceptual dimensions: structured information ($S_t$), residual entropy ($H_t$), observer-dependence, and process reflection. The checklist (presented in full in Appendix~A) offers tiered questions that help designers diagnose where their speculation falls on the quadrant map, identify concrete moves for improving epiplexity, and confirm whether engagement yields transferable structural insight rather than mere affect. It is intended as a reference for design iteration, not a scoring rubric.


\section{Discussion}

\subsection{Reframing quality: from plural criteria to learnable structure}
Our framework does not replace existing speculative quality criteria; rather, it offers a unifying rationale for why certain qualities matter. In Ringfort-Felner et al.'s taxonomy, ``grounded,'' ``reflected,'' and ``participative'' process qualities can be understood as mechanisms for increasing $S_t$: grounding constrains the future-space so causal inference becomes possible; reflection externalizes assumptions and makes learning traceable; participation diversifies interpretive frames so that extracted structures are less parochial \cite{ringfortfelner2025quality,sandersStappers2008cocreation}. Discursive qualities (``experienceable,'' ``thought-provoking'') matter because they shape the channel through which observers extract structure under bounded attention \cite{pirolliCard1999infoForaging}. Speculative qualities (``critical,'' ``socio-political'') matter because the relevant structures of sociotechnical futures are often political and institutional \cite{winner1980artifactsPolitics,jasanoffKim2009sociotechnicalImaginaries}.

The four quadrants provide a more precise vocabulary for common failure modes. The complaint that speculative design produces ``interesting but unclear contribution'' (a familiar peer-review phrase) often describes Q3---high entropy, low epiplexity. The complaint that speculative work is ``incremental'' or ``insufficiently speculative'' describes Q2. The observation that a project had ``great potential but didn't quite land'' may describe Q4---genuine structure that the presentation or engagement context failed to make extractable.

The framework also complements emerging experimental approaches to speculative futures. Rahwan et al.'s science fiction science method provides rigorous tools for testing behavioral responses to speculated technologies, but it does not provide criteria for what makes a speculative scenario \emph{worth testing} \cite{rahwan2025sciFiSci}. Epiplexity fills this gap: it identifies whether a speculation contains learnable structure that justifies the cost of controlled experimentation. A high-epiplexity scenario (Quadrant~I) is a strong candidate for sci-fi-sci investigation because its structured content---second-order effects, governance conditions, value tensions---generates testable hypotheses about human behavior. A low-epiplexity scenario (Quadrant~III) would be a poor investment of experimental resources, no matter how provocative its surface. Conversely, sci-fi-sci provides the empirical validation methods that epiplexity needs: controlled experiments can test whether scenarios designed for high $S_t$ actually produce more transferable learning than those that are not.

Crucially, our framework does \emph{not} impose progressional criteria on frictional work. Pierce argues that frictional design resists progressional evaluation \cite{pierce2021tensionProgression}---and we agree. Epiplexity is \emph{not} a progressional metric: it does not ask ``does this reduce uncertainty toward implementation?'' It asks ``does this yield learnable structure about possibility space?'' Frictional design can have high epiplexity precisely \emph{because} it maintains entropy---but good frictional design makes patterns \emph{inferable} within that maintained uncertainty. The friction is not random; it is strategically placed to reveal structure.

\subsection{Implications for peer review and pedagogy}
Epiplexity suggests practical review questions aligned with speculative design's epistemic aims:
\begin{itemize}
\item \textbf{What structures does the work help us infer?} (Second-order effects, governance conditions, value tensions.)
\item \textbf{Are these structures robust to superficial changes?} (Would insights hold if the artifact's aesthetics or setting changed?)
\item \textbf{What are the bounds of inference?} (What remains uncertain or intentionally open, and why?)
\item \textbf{What is the intermediate-level knowledge claim?} (How is learning made reusable?) \cite{hookLowgren2012strongConcepts,stoltermanWiberg2010conceptDriven}.
\end{itemize}
These questions complement, rather than supplant, existing expectations for reflective accounts in RtD \cite{zimmerman2007rtD,gaver2012expectRtD}.

The four quadrants are directly teachable. A design studio exercise: students analyze existing speculations and place them on the 2$\times$2, using the diagnostic questions to identify why work falls where it does and how to move it. The quadrant labels (``familiar extrapolation,'' ``aestheticized noise,'' ``buried treasure'') provide accessible vocabulary that students can internalize and apply without needing to master information theory. This makes the framework immediately deployable in speculative design pedagogy.

\subsection{Observer-dependence and epistemic pluralism}
The observer-dependence of epiplexity reflects a valuable feature of speculative design: \emph{different audiences need different speculations}. A policymaker anticipating governance frameworks extracts different structure than a designer exploring a problem space or a public reflecting on values. This is not a flaw but a design consideration.

This observer-dependence has implications for participatory and justice-oriented design \cite{costanzaChock2020designJustice,sandersStappers2008cocreation}, as well as for policy-oriented futuring \cite{tsekleves2022speculativePolicy}. If structured information depends on observers' backgrounds, then whose perspectives are centered in speculation matters deeply. Speculative design that achieves high epiplexity for privileged audiences may yield low epiplexity for marginalized communities---and vice versa. The framework formalizes the Prado and Oliveira critique \cite{pradoOliveira2015futuristicGizmos} as an information-theoretic claim: an artifact with high epiplexity only for a narrow audience has genuinely lower epistemic quality than one with high epiplexity across diverse audiences. This reinforces calls for participatory approaches and reflexivity about whose futures are being made legible \cite{bardzell2011feministHCI,philipIraniDourish2012postcolonialComputing}.

\paragraph{Addressing the pluralism-evaluation tension.}
A reasonable concern is that observer-dependence threatens meaningful evaluation: if epiplexity is always relative, how can reviewers make comparative judgments? We propose that legitimate comparison requires:
\begin{enumerate}
\item \textbf{Explicit audience specification.} Authors should state their target observers and engagement conditions. Reviewers then evaluate whether the speculation is well-designed for \emph{those} observers.
\item \textbf{Internal coherence.} Given stated goals, does the artifact's design support the intended learning? A speculation can fail on its own terms.
\item \textbf{Proportionality of claims.} Broader claims require broader transfer. A speculation claiming insight about ``AI futures'' generally should demonstrate robustness across observer types; one claiming insight for ``policymakers in healthcare AI'' can be evaluated more narrowly.
\end{enumerate}
This is not pure relativism---it is contextualized rigor.


\section{Limitations and future work}
This paper offers a \emph{perspectival} rather than formal contribution. We have not empirically validated epiplexity as a measurable construct in speculative design, nor have we provided quantitative operationalizations of $S_t$ and $H_t$. The translation from computational information theory to design contexts is \emph{analogical}: we borrow conceptual structure without claiming mathematical equivalence. This is intentional---we believe the value lies in the new questions the framework prompts---but future work could explore more systematic approaches.

\subsection{Epiplexity and the politics of futures}
An epiplexity lens risks being misread as a search for ``inevitable'' futures, which could slip into technological determinism. We emphasize that ``structure'' in speculative design should include socio-political contingency and contestation. Sociotechnical imaginaries are historically situated and politically constructed \cite{jasanoffKim2009sociotechnicalImaginaries,zuboff2019surveillanceCapitalism}. Feminist and postcolonial HCI remind us that whose futures become legible is itself a design and epistemic question \cite{bardzell2011feministHCI,philipIraniDourish2012postcolonialComputing}. In this sense, increasing $S_t$ can mean increasing the legibility of marginalized structures---racialized harms, labor extraction, accessibility barriers---that dominant imaginaries obscure \cite{benjamin2019raceAfterTechnology,noble2018algorithmsOppression,costanzaChock2020designJustice}.

\subsection{Generative AI as a speculative partner}
Generative AI systems can accelerate ideation and scenario generation, but they also risk homogenizing futures by reproducing dominant narratives and training-data biases \cite{bender2021stochasticParrots,ek2024speculativeAI}. The quadrant framework sharpens this concern: LLMs are adept at generating Quadrant II outputs (bland extrapolations that recombine familiar tropes) and Quadrant III outputs (elaborate, dramatic scenarios that lack structural depth). They struggle with Quadrant I because high $S_t$ requires genuine understanding of causal, institutional, and political structure---not merely plausible-sounding prose.

Epiplexity thus serves as a \emph{steering criterion} for AI-assisted speculation: not ``generate more futures,'' but ``generate futures that maximize learnable structure under bounded attention.'' Designers can use AI to expand the explored future-space in the \emph{explore} phase \cite{cordova2025slrSpeculative}, then apply the audit questions to curate toward Quadrant I---selecting and developing scenarios that surface governance tensions, value conflicts, and institutional dynamics rather than merely novel content.

\paragraph{What would falsify the framework?}
A perspectival contribution resists straightforward falsification, but we can articulate conditions under which the epiplexity lens would prove unhelpful:
\begin{enumerate}
\item Practitioners find the $S_t$/$H_t$ distinction unmappable to their design decisions---they cannot reliably distinguish ``structured insight'' from ``contingent noise'' even with the audit questions.
\item Inter-observer agreement on what counts as structured information proves systematically low, suggesting the distinction is too subjective to guide evaluation.
\item Speculations designed to maximize epiplexity (using the design moves) do not yield more transferable insights than those designed without this lens.
\item The observer-dependence proves so thoroughgoing that no meaningful comparative judgments become possible, collapsing into pure relativism.
\end{enumerate}

\paragraph{Empirical directions.}
We propose four concrete study designs for validation:
\begin{enumerate}
\item \textbf{Controlled comparison.} Hold engagement budget $t$ constant while varying grounding, participation, and discursive format; measure convergence and transferability of inferred structures across observers.
\item \textbf{Inter-observer agreement.} Present the same speculation to diverse observer groups; assess whether they extract similar structural patterns ($S_t$) or whether extracted structures diverge unpredictably.
\item \textbf{Robustness testing.} Generate controlled variations of scenarios (changing superficial details while preserving structural content); measure which insights persist across variations.
\item \textbf{Prospective application.} Have designers explicitly use the epiplexity audit during the design process; compare the learning outcomes of speculations designed with versus without this lens.
\end{enumerate}
Rahwan et al.'s science fiction science method provides a natural experimental framework for these studies \cite{rahwan2025sciFiSci}. Their approach---controlled simulation of speculative futures with systematic measurement of behavioral responses---could be adapted to vary $S_t$ and $H_t$ levels experimentally: present participants with speculations designed for different quadrants and measure whether high-epiplexity scenarios produce more transferable learning (study design~1) and more convergent structural inferences (study design~2). Their fidelity spectrum (text vignettes $\to$ mock applications $\to$ VR $\to$ physical staging) offers a concrete operationalization of the engagement budget $t$, enabling direct tests of how observer resources affect structure extraction.

\paragraph{Temporality and shifting assessments.}
We acknowledge that epiplexity assessments are not fixed across time. What is Q4 (buried treasure) in a gallery setting may become Q1 (structured provocation) in a facilitated workshop with appropriate scaffolding. What is Q1 today may drift to Q2 (familiar extrapolation) as the future it speculates about becomes the present. Future work should explore how the temporal dimension affects framework application.

We also acknowledge that our account foregrounds structured learning; we do not claim that ambiguity, affect, or aesthetic experience are secondary in speculative design. Rather, we argue that for speculative design to sustain its legitimacy in HCI venues, it needs clearer accounts of how such experiences translate into reusable knowledge under bounded conditions \cite{ringfortfelner2025quality,pierce2021tensionProgression}. Epiplexity provides one such account, but others may complement it.


\section{Conclusion}
We proposed an information-theoretic view of speculative design that centers bounded learning from provocation. By adapting epiplexity---structured information extractable by computationally bounded observers \cite{finzi2026epiplexity}---we modeled speculative design as a bounded information process and developed practical tools for evaluation and improvement.

Our theoretical contribution decomposes what observers learn into structured information ($S_t$) and residual entropy ($H_t$), connecting this to emerging quality and process frameworks \cite{cordova2025slrSpeculative,ringfortfelner2025quality} and to convergent findings from complexity science, psychology, and neuroscience about why structured surprise produces learning while noise does not. Our practical contribution---the four-quadrant diagnostic map and the epiplexity audit checklist (Appendix~A)---gives designers, reviewers, and educators a shared vocabulary and a concrete tool for reasoning about speculative design quality.

Crucially, our contribution is \emph{perspectival}: we provide new questions to ask about speculative design (for whom? with what resources? what transfers?) rather than numerical metrics. The observer-dependence of epiplexity is a feature that reflects how speculative design actually works and provides guidance for designers about audience, context, and engagement. We hope this lens provides a practical bridge between speculative design's provocative ambitions and HCI's ongoing need to evaluate and communicate the value of frictional, discursive design research \cite{pierce2021tensionProgression,gaver2012expectRtD}.

\bibliographystyle{ACM-Reference-Format}
\bibliography{reference}

\clearpage
\appendix

\section{Expanded Epiplexity Checklist for Speculative Designers}

This appendix expands the audit questions from Section 4 into a comprehensive reflective checklist, organized by the four conceptual dimensions of the epiplexity framework. It is intended as a reference document for design iteration, not a scoring rubric.

\subsection*{A. Structured Information ($S_t$): Does the speculation yield learnable structure?}

\paragraph{Second-Order Effects and Incentive Structures}
\begin{itemize}
\item Does the speculation surface downstream consequences beyond the immediate technology change?
\item Are incentive structures and behavioral adaptations made inferable?
\item Can observers trace causal chains from the speculated change to social/institutional effects?
\item Does the scenario reveal how metrics, algorithms, or policies reshape behavior over time?
\end{itemize}

\paragraph{Value Tensions and Moral Imaginaries}
\begin{itemize}
\item Does the speculation make competing values visible (e.g., privacy vs.\ convenience, autonomy vs.\ safety)?
\item Are moral trade-offs embedded in the scenario in ways observers can identify?
\item Does the speculation connect to justice concerns (who benefits, who is harmed, who decides)?
\item Can observers articulate the value tensions after engagement, not just feel discomfort?
\end{itemize}

\paragraph{Governance and Contestability Conditions}
\begin{itemize}
\item Does the speculation raise questions about who decides, who audits, who can refuse?
\item Are accountability structures (or their absence) made visible?
\item Can observers infer what institutional arrangements would be needed for legitimacy?
\item Does the scenario make it possible to ask ``who would contest this, and how?''
\end{itemize}

\paragraph{Boundary Conditions and Invariants}
\begin{itemize}
\item Are there constraints that persist across variations of the scenario?
\item Would the insights transfer if we changed superficial details (technology brand, geographic setting, interface style)?
\item Does the speculation reveal something that would matter across multiple plausible futures?
\item Can observers distinguish what is contingent (could be otherwise) from what is structural (likely to persist)?
\end{itemize}

\subsection*{B. Residual Entropy ($H_t$): What is noise, and is it calibrated?}

\paragraph{Identifying Noise}
\begin{itemize}
\item Are there scenario elements that are merely aesthetic or decorative without contributing to inference?
\item Are there contingent details that could be changed without affecting the core insight?
\item Is there complexity that exceeds observers' capacity to interpret given the engagement budget?
\item What would be lost if we removed the most ``striking'' elements---insight or just affect?
\end{itemize}

\paragraph{Checking for Shock Without Structure}
\begin{itemize}
\item If observers react emotionally, can they articulate \emph{why} beyond ``it's disturbing''?
\item Does the provocation direct attention toward specific structural questions, or diffuse it?
\item After the initial surprise fades, what remains to be learned?
\item Is the shock strategically placed to reveal structure, or is it the entire point?
\end{itemize}

\subsection*{C. Observer-Dependence: Who is learning, and under what conditions?}

\paragraph{For Whom?}
\begin{itemize}
\item Who is the intended audience for this speculation?
\item What background knowledge and interpretive frames do they bring?
\item What do we hope \emph{they specifically} will learn?
\item How might different audiences extract different structures from the same artifact?
\end{itemize}

\paragraph{With What Resources?}
\begin{itemize}
\item What is the realistic engagement budget (time, attention, facilitation)?
\item Is the speculation legible within that budget, or does it require extensive explanation?
\item Are there scaffolds (documentation, workshops, props, guided discussion) to support interpretation?
\item What would observers miss if they had half the engagement time?
\end{itemize}

\paragraph{What Transfers?}
\begin{itemize}
\item Can insights from this speculation inform decisions beyond this specific project?
\item Will the learnings matter in six months? Two years?
\item Can the structured information be articulated as intermediate-level knowledge (strong concepts, design considerations, policy questions)?
\item Would a different design team, encountering this speculation, extract similar structures?
\end{itemize}

\subsection*{D. Process Reflection: How does the design process affect epiplexity?}

\paragraph{Grounding}
\begin{itemize}
\item Is the speculation connected to plausible trajectories, real constraints, or lived realities?
\item Are the assumptions explicit and contestable?
\item What research (empirical, historical, technical) informs the scenario?
\end{itemize}

\paragraph{Participation}
\begin{itemize}
\item Whose perspectives shaped the speculation?
\item Are marginalized structures (labor conditions, racialized harm, accessibility barriers) made legible?
\item Did diverse observers help test whether structure is extractable?
\end{itemize}

\paragraph{Reflexivity}
\begin{itemize}
\item What are our own biases as speculative designers?
\item What futures did we \emph{not} explore, and why?
\item How would different positionalities produce different speculations---and different epiplexity?
\end{itemize}

\vspace{1em}
\noindent This checklist is not a scoring rubric; it is a set of prompts for reflection and design iteration. High-quality speculation need not answer ``yes'' to every question, but engaging with these questions can help designers craft provocations that yield learnable structure rather than mere shock.

\section{Disclosure of the Usage of LLM}
We used Claude (Claude 4.5 Opus model) to assist with writing this manuscript. Specific uses included:
\begin{itemize}
\item Brainstorming and refining the conceptual framework
\item Literature review and reference verification
\item Drafting and editing prose
\item Developing the Epiplexity Checklist and audit questions
\end{itemize}

\end{document}